\documentclass[aps,showpacs,amsmath,amssymb,twocolumn,pra,superscriptaddress,notitlepage]{revtex4-2}

\usepackage{amsmath,bm}
\usepackage[dvipdfmx]{graphicx}
\usepackage{amsmath,amssymb,amsthm,mathrsfs,amsfonts,dsfont}
\usepackage{braket}
\usepackage{color}
\usepackage{graphicx}
\usepackage[linesnumbered, ruled,vlined]{algorithm2e}
\usepackage{braket}
\usepackage{comment}
\usepackage{appendix}
\usepackage{here}
\usepackage{tabularx}
\usepackage{multirow}
\usepackage{mathtools}
\usepackage{qcircuit}
\usepackage{enumerate}
\usepackage{subfigure, epsfig}

\DeclarePairedDelimiter\floor{\lfloor}{\rfloor}
\newcolumntype{Y}{>{\centering\arraybackslash}X}

\begin{document} 


\title{Resource estimations for the Hamiltonian simulation in correlated electron materials}


\author{Shu Kanno}
\email{kanno.s.ac@m.titech.ac.jp}
\affiliation{Department of Materials Science and Engineering, Tokyo Institute of Technology, 2-12-1 O-okayama, Meguro-ku, Tokyo 152-8552, Japan}

\author{Suguru Endo}
\affiliation{NTT Computer and Data Science Laboratories, NTT Corporation, Musashino 180-8585, Japan}
\affiliation{JST, PRESTO, 4-1-8 Honcho, Kawaguchi, Saitama, 332-0012, Japan}

\author{Takeru Utsumi}
\affiliation{Department of Materials Science and Engineering, Tokyo Institute of Technology, 2-12-1 O-okayama, Meguro-ku, Tokyo 152-8552, Japan}

\author{Tomofumi Tada}
\affiliation{Department of Materials Science and Engineering, Tokyo Institute of Technology, 2-12-1 O-okayama, Meguro-ku, Tokyo 152-8552, Japan}
\affiliation{Materials Research Center for Element Strategy, Tokyo Institute of Technology, 4259 Nagatsuta, Midori-ku, Yokohama, Kanagawa 226-8501, Japan}
\affiliation{Kyushu University Platform of Inter/Transdisciplinary Energy Research (Q-PIT), Kyushu University, Fukuoka 819-0395 Japan}


\begin{abstract}
Correlated electron materials, such as superconductors and magnetic materials, are regarded as fascinating targets in quantum computing. 
However, the quantitative resources, specifically the number of quantum gates and qubits, required to perform a quantum algorithm to simulate correlated electron materials remain unclear.
In this study, we estimate the resources required for the Hamiltonian simulation algorithm for correlated electron materials, specifically for organic superconductors, iron-based superconductors, binary transition metal oxides, and perovskite oxides, using the fermionic swap network. 
The effective Hamiltonian derived using the $ab~initio$ downfolding method is adopted for the Hamiltonian simulation, and a procedure for the resource estimation by using the fermionic swap network for the effective Hamiltonians including the exchange interactions is proposed.
For example, in the system for the $10^2$ unit cells, the estimated number of gates per Trotter step and qubits are approximately $10^7$ and $10^3$, respectively, on average for the correlated electron materials.
Furthermore, our results show that the number of interaction terms in the effective Hamiltonian, especially for the Coulomb interaction terms, is dominant in the gate resources when the number of unit cells constituting the whole system is up to $10^2$, whereas the number of fermionic swap operations is dominant when the number of unit cells is more than $10^3$.
\end{abstract}

\maketitle

\section{introduction}
\label{sec:introduction}
Quantum computers are expected to more accurately solve computational problems in quantum chemistry and materials science beyond the conventional computers~\cite{Cao2019-qa,McArdle2020-fz,Bauer2020-ug,Moll2018-xx,Elfving2020-oe,Motta2021-cr,Arute2019-hj,Zhong2020-bf,Wu2021-wj,Zhong2021-zb,Endo2021-ku,Cerezo2020-un,Tilly2021-pb,Huang2022-ke}.
One key algorithm in quantum computation is the Hamiltonian simulation algorithm~\cite{Feynman1982-gy,Lloyd1996-wk}, which simulates the real-time evolution of quantum systems. 
The Hamiltonian simulation can be used also for calculating static properties of the system using the quantum phase estimation algorithm~\cite{Aspuru-Guzik2005-vu,Yu_Kitaev1995-cf,Nielsen2010-hw}. The experimental demonstrations for the Hamiltonian simulation of lattice models~\cite{Lanyon2011-ww,Georgescu2014-uk,Arute2020-fy} and small molecules~\cite{Lu2011-ij,Sparrow2018-zv} have been performed using quantum computers.

For applications on the computational problems in the future, resource estimations such as the number of qubits and quantum gates to perform the Hamiltonian simulation are important to mark a milestone. Hereafter, we represent the number of qubits and quantum gates as qubit and gate resources, respectively.
Previously, resource estimations have been conducted for molecules~\cite{Wecker2014-va,Reiher2017-jt}, a jellium~\cite{Kivlichan2020-sp,McArdle2022-tz}, and the Hubbard model~\cite{Clinton2021-jf}. 
However, the resource estimation for practical correlated electron materials such as unconventional superconductors, magnetic materials, and Mott insulators has not been reported even though research on correlated materials is a central topic in materials science~\cite{Bauer2020-ug,Imada2010-pb,Kotliar2004-ht}.

In this work, we estimate the qubit and gate resources required for correlated electron materials in the Trotter-based Hamiltonian simulation. 
We consider the resource estimation for the single Trotter step of the Hamiltonian simulation using the fermionic swap (fswap) network~\cite{Kivlichan2018-gz,Kivlichan2020-sp,Cai2020-sv,Hagge2020-ba,Babbush2018-qp} , where the fswap network is a method for estimating resources in quantum devices with the nearest-neighbor connectivity (e.g., superconducting and silicon qubit devices).
The estimation has been conducted for 13 compounds of organic superconductors~\cite{Jerome1991-tt}, iron-based superconductors 
~\cite{Kamihara2006-bu,Kamihara2008-ty}, binary transition metal oxides $TM$O ($TM$ = Mn, Fe, Co, Ni)~\cite{Zaanen1985-zc,Shih2012-df,Sakuma2013-za}, and perovskite oxides $\mathrm{Sr}M\mathrm{O_3}$ ($M$ = V, Cr, Mn)~\cite{Pena2001-gc,Vaugier2012-th}. The organic and iron-based superconductors are known as unconventional superconductors~\cite{Nakamura2016-vj,Nomura2012-um,Miyake2010-ml}. The binary transition metal oxides are Mott or charge-transfer insulators~\cite{Shih2012-df,Sakuma2013-za}, and some perovskite oxides show magnetic properties~\cite{Vaugier2012-th,Imada2010-pb,Imada1998-je}.
In this study, we adopt effective Hamiltonians with orbitals around the Fermi level that mainly contribute to the electron correlation phenomena in the compounds (e.g., Fe $d$-orbitals of the iron-based superconducting material) based on the classical electronic structure calculation method, called the $ab~initio$ downfolding \cite{Imada2010-pb}. 
The qubit resource is determined depending on the size of the effective Hamiltonian. 
For estimating the gate resource, we extend the fswap network to the effective Hamiltonian including the operations on four spin-orbitals because the effective Hamiltonian in this study is based on the localized basis, and the interactions on the four spin-orbitals are necessary; the fswap network in the previous works~\cite{Kivlichan2018-gz,Cai2020-sv,Kivlichan2020-sp} is limited to the interaction up to two spin-orbitals, and thus the methods of the previous study cannot be directly applied to our study.
Notably, this approach can be straightforwardly applied to (Trotter-based) near-term algorithms, such as the variational quantum algorithm using the Hamiltonian variational ansatz~\cite{Peruzzo2014-kp,Wecker2015-fu}.

The rest of this paper is organized as follows.
First, an overview of the Trotter-based Hamiltonian simulation is described in Sec.~\ref{sec:overview of Hs}. The procedures for obtaining the effective Hamiltonians and fswap network are respectively explained in Sec.~\ref{sec:Construction of effective Hamiltonian} and Sec.~\ref{sec:Fermionic swap network and the procedure in the Hamiltonian with the exchange interaction}. The results of resource estimation are presented in Sec.~\ref{sec:Gate resource estimation results}, and future directions of this study are discussed in Sec.~\ref{sec:Conclusion}.

\section{Results}
\label{sec:results}
\subsection{Trotter-based Hamiltonian simulation}
\label{sec:overview of Hs}
Assuming the Hamiltonian $H$ of the system is represented in terms of the local Hamiltonian, which acts on a small subset of the spin-orbitals, we can write the Hamiltonian as
\begin{equation}
\begin{aligned}
H &=\sum_i \hat{h}_i,
\label{Eq:Hamiltonian decomposition}
\end{aligned}
\end{equation}
where $\hat{h}_i$ is the local Hamiltonian. Due to the first order Trotter formula, the ideal unitary operator of time-evolution for some duration $t$, $U_{ideal}=e^{-i H t}$, can be approximated as
\begin{equation}
\begin{aligned}
U_{ideal}&=U_{Trotter}^{\frac{t}{\Delta t}}+ O(t \Delta t),
\label{Eq:unitary hs}
\end{aligned}
\end{equation}
where $U_{Trotter}=\prod_i e^{-i \hat{h}_i \Delta t }$, and $\Delta t$ is a single time step. Therefore, the real time dynamics can only be simulated using the local operations.
Henceforth, we will refer to $e^{-i \hat{h}_i \Delta t }$ in $U_{Trotter}$ as the interaction operator.

We estimate the gate resources required for $U_{Trotter}$, i.e., the single Trotter step,
by evaluating the required number of controlled-NOT (CNOT) and arbitrary single-qubit gates for performing each interaction operator, whereas the qubit resource is determined by the Hamiltonian size since the qubit resource corresponds to the number of spin-orbitals in the Hamiltonian.

\begin{table*}[!ht]
\begin{center}
\caption{Classification of the compounds, target orbital, and number of qubits (spin-orbitals) per unit cell $N_{qubits/cell}$. TMTSF denotes tetramethyltetraselenafulvalene, and $\mathrm{K_3C_{60}}$ is called K-doped fullerene. Doping and/or external pressure are required to experimentally observe the superconductivity for LaFeAsO and $\mathrm{BaFe_2As_2}$ in the iron-based superconductor class and $\mathrm{(TMTSF)_2 PF_6}$ in the organic superconductor class.}
\label{tab:compounds list}
    \begin{tabular}{c c c c}
    \hline
        Classification & Compound & Target orbital & $N_{qubits/cell}$ \\ \hline
        \multirow{2}{*}{Organic superconductor} & $\mathrm{(TMTSF)_2PF_6}$ & Linear combination of $p$-orbitals in TMTSF molecule$\times$2 & 4 \\ 
        ~ & $\mathrm{K_3C_{60}}$ & $p$-orbitals at a fullerene cage & 6 \\ \hline
        \multirow{4}{*}{Iron-based superconductor} & LaFeAsO & $d$-orbitals in Fe$\times$2 & 20 \\ 
        ~ & $\mathrm{BaFe_2As_2}$ & $d$-orbitals in Fe$\times$2 & 20 \\ 
        ~ & LiFeAs & $d$-orbitals in Fe$\times$2 & 20 \\ 
        ~ & FeSe & $d$-orbitals in Fe$\times$2 & 20 \\ \hline
        \multirow{4}{*}{Binary transition metal oxide} & MnO & $d$-orbitals in Mn & 10 \\ 
        ~ & FeO & $d$-orbitals in Fe & 10 \\ 
        ~ & CoO & $d$-orbitals in Co & 10 \\ 
        ~ & NiO & $d$-orbitals in Ni & 10 \\ \hline
        \multirow{3}{*}{Perovskite oxide} & $\mathrm{SrVO_3}$ & $d$-orbitals in V & 10 \\ 
        ~ & $\mathrm{SrCrO_3}$ & $d$-orbitals in Cr & 10 \\ 
        ~ & $\mathrm{SrMnO_3}$ & $d$-orbitals in Mn & 10 \\ \hline
    \end{tabular}

\end{center}
\end{table*}

\subsection{Construction of the effective Hamiltonian}
\label{sec:Construction of effective Hamiltonian}
The effective Hamiltonian $H$ is defined as
\begin{equation}
\begin{aligned}
H&=\sum_{\sigma} \sum_{p} t_{p p \sigma} n_{p \sigma}+\sum_{\sigma} \sum_{p<q} t_{p q \sigma}(a_{p \sigma}^{\dagger} a_{q \sigma}+a_{p \sigma} a_{q \sigma}^{\dagger})\\
&+\sum_{p} U_{p p} n_{p \uparrow} n_{p \downarrow}+\sum_{\sigma, \sigma^{\prime}} \sum_{p<q} U_{p q} n_{p \sigma} n_{q \sigma^{\prime}}\\
&-\sum_{\sigma} \sum_{p<q} J_{p q} n_{p \sigma} n_{q \sigma} \\
&+\sum_{p<q} J_{p q} (a_{p \uparrow}^{\dagger} a_{p \downarrow} a_{q \uparrow} a_{q \downarrow}^{\dagger}+a_{q \downarrow} a_{q \uparrow}^{\dagger} a_{p \downarrow}^{\dagger}  a_{p \uparrow})\\
&-\sum_{p<q} J_{p q} (a_{p \uparrow}^{\dagger} a_{p \downarrow}^{\dagger} a_{q \uparrow} a_{q \downarrow} + a_{q \downarrow}^{\dagger} a_{q \uparrow}^{\dagger} a_{p \downarrow} a_{p \uparrow}),
\label{eq:effective H}
\end{aligned}
\end{equation}
where the first and second terms come from one-body potentials, the third and fourth terms follow from the Coulomb interactions, and the last three terms from exchange interactions.
$\sigma$ and $\sigma'$ are the spin indices, $p$ and $q$ are the orbital indices, 
$a^{\dag}_{p\sigma}$ ($a_{p\sigma}$) is the creation (annihilation) operator on spin-orbitals of $p\sigma$, and $n_{p\sigma}$ is $a_{p\sigma}^{\dag}a_{p\sigma}$ and is called the particle number operator. $t_{pq\sigma}$ is the hopping integral between spin-orbitals of $p\sigma$ and $q\sigma$, and $U_{pq}$ and $J_{pq}$ are the effective Coulomb and effective exchange interactions between the orbitals of $p$ and $q$, respectively. 
Equation~(\ref{eq:effective H}) is a general form of the effective Hamiltonian in the localized basis, and please see Ref.~\cite{Misawa2019-vt} for details of the Hamiltonian. 
Note that we arranged the index sequence of the last two terms in Eq.~(\ref{eq:effective H}) as $p\uparrow, p\downarrow, q\uparrow,$ and $q\downarrow$ from that in Ref.~\cite{Misawa2019-vt} for the swapping operation described later. 
We denote an initial sequence of qubit indices for the spin-orbital indices $p\sigma$ in the circuit by $(1\uparrow, 1\downarrow, 2\uparrow, 2\downarrow, \dots)$ and use Jordan-Wigner encoding~\cite{Jordan1928-hf} to transform the fermionic operators $a^{\dag}_{p\sigma}$ and $a_{p\sigma}$ to the Pauli operators. Furthermore, we consider systems under the periodic boundary condition and including only static interactions (zero frequency) in the Hamiltonian.
Hereafter, we will refer to $t_{pp\sigma}$ and $t_{pq\sigma}$ as $t$, $U_{pp}$ and $U_{pq}$ as $U$, and $J_{pq}$ as $J$.

Table~\ref{tab:compounds list} shows the list of the target compounds.
The compounds are classified as organic superconductors, iron-based superconductors, binary transition-metal oxides $TM$O ($TM$ = Mn, Fe, Co, Ni), and perovskite oxides $\mathrm{Sr}M\mathrm{O_3}$ ($M$ = V, Cr, Mn).
For all the target compounds, we determined the crystal structure and target orbitals according to the previous studies~\cite{Nakamura2016-vj,Nomura2012-um,Miyake2010-ml,Shih2012-df,Sakuma2013-za,Vaugier2012-th} by considering the condition that orbitals appearing in the effective Hamiltonian should be close to the Fermi level and mainly contribute to the electron correlation phenomena.

The interaction coefficients $t$, $U$, and $J$ in Eq.~(\ref{eq:effective H}) were calculated using the $ab~initio$ downfolding method, consisting of three steps, namely, the band-structure calculation for the target material, calculation of the target orbitals and $t$, and calculation of $U$ and $J$ (the details of the procedures are described elsewhere~\cite{Imada2010-pb,Aryasetiawan2004-tc,Misawa2012-xn,Miyake2010-ml,Misawa2011-cw,Ohgoe2020-fb,Kanno2021-no,Kanno2021-rg}).
First, we calculated the band-structure using density functional theory (DFT) in a non-spin-polarized calculation, as implemented in Quantum ESPRESSO package~\cite{Giannozzi2009-zq,Giannozzi2017-jz,Giannozzi2020-yi}. 
We adopted the generalized gradient approximation by Perdew-Burke-Ernzerhof as the exchange-correlation functional~\cite{Perdew1996-qc} and the norm-conserving pseudopotential~\cite{Hamann1979-ti,Hamann2013-tx}. 
Second, the orbitals of the effective Hamiltonian were obtained using the maximally localized Wannier function~\cite{Marzari1997-sz} (hereafter Wannier orbital) with the target orbitals in Table~\ref{tab:compounds list}. 
$t_{pq\sigma}$ (and $t_{pp\sigma}$) is obtained from the Wannier orbitals as
\begin{equation}
    \begin{aligned}
    t_{pq\sigma} = \int \psi_p^*(\textbf{r}) H_0 \psi_q(\textbf{r}) d\textbf{r},
    \label{eq:transfer integral}
    \end{aligned}
\end{equation}
where $\psi_p(\textbf{r})$ is the $p$-th Wannier orbital, and $H_0$ is the Kohn-Sham Hamiltonian calculated using the DFT. The integral is taken over the crystal volume.
Third, $U_{pq}$ (and $U_{pp}$) and $J_{pq}$ are obtained as
\begin{equation}
    \begin{aligned}
    U_{pq} = \iint |\psi_p(\textbf{r})|^2 W(\textbf{r},\textbf{r}') |\psi_q(\textbf{r}')|^2 d\textbf{r}d\textbf{r}'
    \label{eq:effective coulomb}
    \end{aligned}
\end{equation}
\begin{equation}
    \begin{aligned}
    J_{pq} = \iint \psi_p^*(\textbf{r})\psi_q(\textbf{r}) W(\textbf{r},\textbf{r}') \psi_q^*(\textbf{r}') \psi_p(\textbf{r}') d\textbf{r}d\textbf{r}',
    \label{eq:effective coulomb}
    \end{aligned}
\end{equation}
where $W(\textbf{r},\textbf{r}')$ is the screened Coulomb interaction calculated using the constrained random phase approximation~\cite{Aryasetiawan2004-tc}. 
The second and third steps were performed using RESPACK package~\cite{Fujiwara2003-xg,Nakamura2008-hg,Nakamura2009-te,Nohara2009-rm,Nakamura2016-vj,Nakamura2021-sh}. 
The convergence parameters (the wave-function cutoff, polarization-function cutoff, $k$-point grids, and the number of bands) in the procedures were determined to maintain that the averaged $U_{pp}$ converged within 0.1 eV.
We adopted threshold values of $|t|$, $U$, and $J$ as 0.01, 0.20, and 0.20 eV, respectively, since small values for the interaction coefficients do not affect computational results~\cite{Ohgoe2020-fb,Misawa2012-xn}. Additionally, we carefully considered the $k$-point grids by the computational condition in RESPACK~\footnote{As a computational condition in RESPACK, we confirmed the adoption of $k$-point grids sufficient to obtain all the interactions above the threshold of $|t|$, $U$, and $J$ because the maximum distance of the calculated interaction depends on the $k$-point grids.}.

We define the number of unit cells, $N_{cells}$, as a variable to estimate the resources because the simulations with larger sizes than the unit cell will be performed in future applications (e.g., nonequilibrium dynamical simulations). 
Thus, the qubit resource is determined by multiplying the number of qubits per unit cell (``$N_{qubit/cell}$'' in Table~\ref{tab:compounds list}) with $N_{cells}$ and is about one order of magnitude larger than $N_{cells}$ on average for the target compounds. For example, in LaFeAsO for $10^2$ unit cells (e.g., 2D $10\times10$ system), $2\times10^3$ qubits are required. 
Note that the quantum device of the scale of $10^3$ qubits may appear in the near future~\cite{Matthews2021-pl}.
We will mainly discuss the gate resource estimation in the rest of this paper.

\subsection{Fermionic swap network and the procedure in the Hamiltonian with the exchange interactions}
\label{sec:Fermionic swap network and the procedure in the Hamiltonian with the exchange interaction}
When the Hamiltonian includes long-range interactions, the interaction operations between distant qubits in Eq.~(\ref{Eq:unitary hs}) are required to perform the Hamiltonian simulation.
However, the operations are not directly executable on devices in which the operations are restricted to the neighboring qubit, such as superconducting and silicon qubit devices.
A fswap network~\cite{Kivlichan2018-gz} is a method that can perform the Hamiltonian simulations using only the nearest-neighbor operations, which are accomplished by the swapping operation between spin-orbital indices, the fswap operation. The fswap operator between two indices $p\sigma$ and $q\sigma'$ is defined as $f_\mathrm{swap}^{p\sigma, q\sigma'}=1 + a^{\dag}_{p\sigma} a_{q\sigma'} + a^{\dag}_{q\sigma'} a_{p\sigma} - a^{\dag}_{p\sigma} a_{p\sigma} - a^{\dag}_{q\sigma'} a_{q\sigma'}$ and holds properties of $f_\mathrm{swap}^{p\sigma, q\sigma'} a^{\dag}_{p\sigma} (f_\mathrm{swap}^{p\sigma, q\sigma' })^{\dag} = a^{\dag}_{q\sigma'}$ and $f_\mathrm{swap}^{p\sigma, q\sigma'} a_{p\sigma} (f_\mathrm{swap}^{p\sigma, q\sigma'})^{\dag} = a_{q\sigma'}$. 
For example, in a device with the one-dimensional and nearest-neighbor qubit coupling, the fswap network in the Hamiltonian simulation using one and two spin-orbitals has been implemented by executing the interaction and fswap operations between the neighboring qubits in a simply interchangeable manner~\cite{Kivlichan2018-gz}.
The fswap network has also been applied to several fermionic Hamiltonians (including only interactions of one or two spin-orbitals) related to quantum chemistry~\cite{Kivlichan2020-sp,Babbush2018-qp} and the Hubbard model~\cite{Cai2020-sv,Hagge2020-ba}.
In this study, we propose the procedure of the fswap network to the Hamiltonian with the exchange interaction terms that operate on four spin-orbitals (see the last two terms in Eq.~(\ref{eq:effective H})).
We mention that a procedure for executing the fswap network on the interaction operators of four spin-orbitals was proposed in the context of the unitary coupled cluster~\cite{OGorman2019-eh}, although the number of fswap operators is larger than that of our procedure by $\frac{N_{qubits/cell}N_{cells}}{2}$.

\begin{figure}[!ht]
 \includegraphics[width=1\columnwidth]{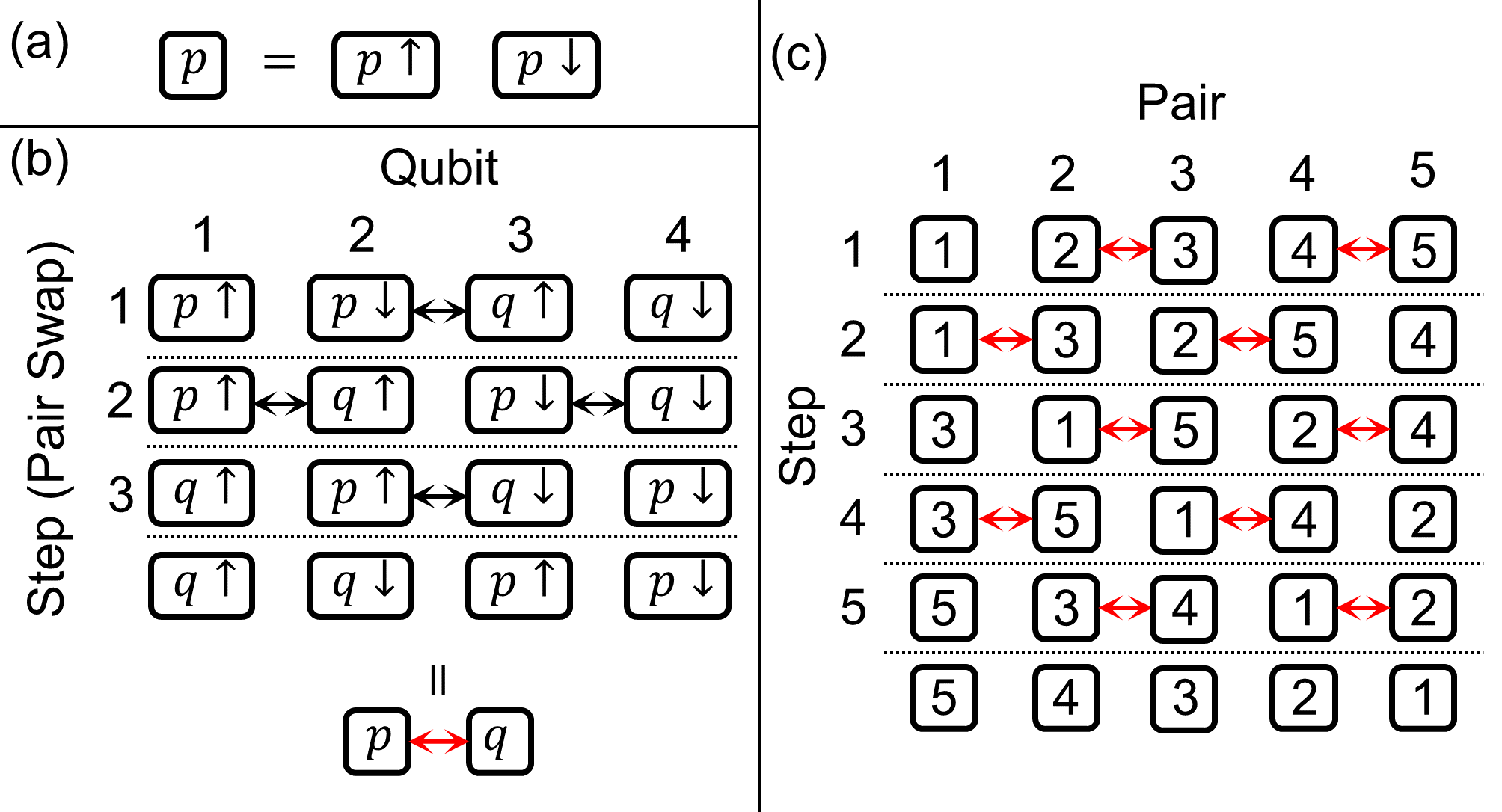}
\caption{Procedure of the fswap network for a one-dimensional array of 10 qubits (= 10 spin-orbitals, 5 orbitals). $p$ and $p\sigma$ ($q\sigma$) in the boxes are the indices of the orbital and spin-orbital, respectively, where $p, q \in [1,5]$ and $\sigma = \uparrow, \downarrow$. (a) Representation of the pair of the spin-orbitals. (b) The pair swapping operation. (c) Depiction of the procedure.}
 \label{fig:fswap_procedure_1col}
\end{figure}

Figure~\ref{fig:fswap_procedure_1col} shows the fswap network procedure in a device with the one-dimensional qubit coupling for the Hamiltonian simulations including the operations on four spin-orbitals.
Here, the fswap operation is executed on pairs of up and down spins with the same orbital index as in Fig.~\ref{fig:fswap_procedure_1col}(a).
The main component of the procedure is the swapping operation between the pairs specified by two orbital indices (pair swapping operation), as in Fig.~\ref{fig:fswap_procedure_1col}(b).
At each step, we interchangeably execute the interaction operations executable in the spin-orbital sequence and fswap operations between spin-orbitals on both sides of the black double-headed arrows.
The four fswap operations are used in the pair swapping operation in Fig.~\ref{fig:fswap_procedure_1col}(b).
The interaction and fswap operators are implemented by CNOT and arbitrary one-qubit gates (see Appendix~\ref{The quantum circuit implementations in the fswap network} for the circuit implementations).

The whole fswap network procedure is depicted in Fig.~\ref{fig:fswap_procedure_1col}(c), and its pseudocode is given as Algorithm~\ref{Alg:fswap}. 
At each step in Fig.~\ref{fig:fswap_procedure_1col}(c), the pair swapping operations between the two pairs of the orbital index marked with red arrows are executed.
The red arrows exist between $2j+2$ and $2j+3$ pairs in the odd step and $2j+1$ and $2j+2$ in the even step, where $j \in [0, \floor{\frac{N_o-2}{2}}]$, and $N_o$ is the number of pairs.
The operations are executed until the sequence of the orbital indices is reversed, as shown at the bottom in Fig.~\ref{fig:fswap_procedure_1col}(c).
Since the four indices of $p\uparrow, p\downarrow, q\uparrow,$ and $q\downarrow$ appear in this sequence for any $p$ and $q$ at some step during the fswap network, the procedure can be applied to the Hamiltonian with the terms for the four spin-orbitals as in exchange interaction terms.

\begin{algorithm}[!b]
\caption{Fswap network in the effective Hamiltonian with the exchange interactions}
\label{Alg:fswap}
\SetKwInput{KwInput}{Input}                
\SetKwInput{KwOutput}{Output}              
\SetKwInOut{Parameter}{Global variable}
\SetKwFunction{FMain}{Main}
\SetKwFunction{PairSwap}{PairSwap}
\SetKwProg{Fn}{Function}{:}{}
\SetKwRepeat{Do}{do}{while}

\DontPrintSemicolon
  
  \KwInput{WFini (Initial wave function), IntList (List of the interaction operators), $N_o$ (The number of orbitals)}
  \KwOutput{WF (Wave function after the single Trotter step, and the sequence of the orbital indices is reversed from that of the initial wave function)}
  \Parameter{ExecutedList, WF}
  ExecutedList $\gets$ Empty list\;
  WF $\gets$ WFini
  \;

  \Fn{\PairSwap{$p$, $q$}}{
        \For{$step\_ps = 1, 2, 3$}{
        TempIntList $\gets$ Empty list\;
        TempIntList $\gets$ The interaction operators in IntList that are executable in the current spin-orbital sequence and not in ExecutedList\;
        Execute the interaction operations in TempIntList on WF \;
        Add the executed interaction operators of TempIntList to ExecutedList \;
            \If{$step\_ps = 1$} {
                Execute $f_\mathrm{swap}^{p\downarrow, q\uparrow}$ on WF\;
            }
            \If{$step\_ps = 2$} {
                Execute $f_\mathrm{swap}^{p\uparrow, q\uparrow}$ and $f_\mathrm{swap}^{p\downarrow, q\downarrow}$ on WF\;
            }
            \If{$step\_ps = 3$} {
                Exeute $f_\mathrm{swap}^{p\uparrow, q\downarrow}$ on WF\;
            }
        }
  }

  \Fn{\FMain}{
        $step$ $\gets$ 1\;
        \While{true}{
            \If{$step$ is odd} {
                \For{$j =0$ to $\floor{\frac{N_o-2}{2}}$}{
                        \textbf{if} $2j+3>N_o$ \textbf{then}; \textbf{break}\;
                        $p$ $\gets$ orbital index of $2j+2$ pair\;
                        $q$ $\gets$ orbital index of $2j+3$ pair\;
                        PairSwap($p$, $q$)
                    }
            }
            \If{$step$ is even} {
                \For{$j =0$ to $\floor{\frac{N_o-2}{2}}$}{
                        $p$ $\gets$ orbital index of $2j+1$ pair\;
                        $q$ $\gets$ orbital index of $2j+2$ pair\;
                        PairSwap($p$, $q$)\;
                    }
            }
            \If{the sequence of the orbital indices coincides with the reversed sequence from the initial one}{
            \textbf{break}
            }
        $step$ $\gets$ $step+1$\;
        }
        \Return WF\;
  }
\end{algorithm}

Now, we explain the detail of the gate resource estimation using the fswap network based on CNOT gates and arbitrary one-qubit gates.
We estimate the number of gates required for the single Trotter step $U_{Trotter}$ in Eq.~(\ref{Eq:unitary hs}).
The contributions to the gate resources for the interaction and the fswap operations can be described separately.  
The gate resource $N_{gates}$ is represented as
\begin{equation}
\begin{aligned}
N_{gates}&=\sum_{i_{int}} N_{terms}^{i_{int}} N_{gates}^{i_{int}}\\
&+ N_{pairswap} (4 N_{gates\mathchar`-fswap}),\\
\label{eq:Ngates}
\end{aligned}
\end{equation}
where the first and second terms in the right side expression represent the gate resources for the interaction and fswap operators, respectively.
$i_{int}$ is the interaction index, i.e., corresponds to one of the seven interactions in Eq.~(\ref{eq:effective H}) (e.g., $t_{p p \sigma} n_{p \sigma}$). 
$N_{terms}^{i_{int}}$ is the number of interaction operators (or terms in the Hamiltonian) specified by $i_{int}$; $N_{gates}^{i_{int}}$ is the number of gates required for an interaction operator specified by $i_{int}$; $N_{pairswap}$ is the number of the pair swapping operations; $N_{gates\mathchar`-fswap}$ is the number of gates required for the single fswap operation.
The values of $N_{gates}^{i_{int}}$ are listed in Table~\ref{tab:the number of gates}. We considered three cases where one value for the one-qubit gate, that of the CNOT gate, or the summation of the values is assigned to $N_{gates}^{i_{int}}$.
Also, the value of $N_{gates\mathchar`-fswap}$ in the one-qubit gate, that of the CNOT gate, and the summation of the values are $2, 2$, and $4$, respectively.
Since Eq.~(\ref{eq:Ngates}) is an expression with the system size $N_{cells}$ implicitly included in each term, we transformed Eq.~(\ref{eq:Ngates}) into the expression that explicitly includes $N_{cells}$ for estimating the resource as a function of $N_{cells}$,
\begin{equation}
\begin{aligned}
N_{gates}[N_{cells}]&=(\sum_{i_{int}} N_{terms/cell}^{i_{int}} N_{gates}^{i_{int}}) N_{cells}\\
&+\frac{1}{2}(N_{qubits/cell}^2N_{cells}^2\\
&-2N_{qubits/cell}N_{cells}) N_{gates\mathchar`-fswap},
\label{eq:Ngates(Ncell)}
\end{aligned}
\end{equation}
where $N_{terms/cell}^{i_{cell}}$ is the value of $N_{terms}^{i_{int}}$ per unit cell. 
We used the relation $N_{pairswap}={\frac{N_{qubits/cell} N_{cells}}{2} \choose 2}$ to obtain Eq.~(\ref{eq:Ngates(Ncell)}) from Eq.~(\ref{eq:Ngates}).
We estimated the gate resource using Eq.~(\ref{eq:Ngates(Ncell)}).
 
\begin{table}[!b]
\begin{center}
\caption{The number of gates, $N_{gates}^{i_{int}}$, required for implementing each interaction operator of $i_{int}$. We showed the values of one-qubit gates, CNOT gates, and the summation of the values.}
\label{tab:the number of gates}

    \begin{tabular}{c c c c}
    \hline
        \multirow{2}{*}{Interaction index ($i_{int}$)} & \multicolumn{3}{c}{$N_{gates}^{i_{int}}$} \\ \cline{2-4}
          & 1qubit & CNOT & Sum \\ \hline
        $t_{p p \sigma} n_{p \sigma}$ & 1 & 0 & 1 \\ 
        $t_{p q \sigma}(a_{p \sigma}^{\dagger} a_{q \sigma}+a_{p \sigma} a_{q \sigma}^{\dagger})$ & 10 & 4 & 14 \\ 
        $U_{p p} n_{p \uparrow} n_{p \downarrow}$ & 4 & 2 & 6 \\ 
        $U_{p q} n_{p \sigma} n_{q \sigma^{\prime}}$ & 4 & 2 & 6 \\ 
        $J_{p q} n_{p \sigma} n_{q \sigma}$ & 4 & 2 & 6 \\ 
        $J_{p q} (a_{p \uparrow}^{\dagger} a_{p \downarrow} a_{q \uparrow} a_{q \downarrow}^{\dagger}+a_{q \downarrow} a_{q \uparrow}^{\dagger} a_{p \downarrow}^{\dagger}  a_{p \uparrow})$ & 72 & 48 & 120\\
        $J_{p q} (a_{p \uparrow}^{\dagger} a_{p \downarrow}^{\dagger} a_{q \uparrow} a_{q \downarrow} + a_{q \downarrow}^{\dagger} a_{q \uparrow}^{\dagger} a_{p \downarrow} a_{p \uparrow})$ & 72 & 48 & 120 \\ \hline
    \end{tabular}
\end{center}
\end{table}

\begin{table*}[!ht]
\begin{center}
\caption{List of the estimated results for the number of gates in the target compounds.}
\label{tab:List of the resource estimation results}
    \begin{tabularx}{\textwidth}{c c Y Y Y Y Y Y Y Y Y}
    \hline
        \multirow{2}{*}{Classification} & \multirow{2}{*}{Compound} & \multicolumn{3}{c}{$N_{gates}[N_{cells}=10^2]$} & \multicolumn{3}{c}{$N_{gates}[N_{cells}=10^3]$} & \multicolumn{3}{c}{$N_{gates}[N_{cells}=10^4]$} \\ \cline{3-11}
        ~ & ~ & 1qubit & CNOT & Sum & 1qubit & CNOT & Sum & 1qubit & CNOT & Sum \\ \hline 
        \multirow{2}{*}{}Organic & $\mathrm{(TMTSF)_2PF_6}$ & $3.9\times10^{5}$ & $2.7\times10^{5}$ & $6.7\times10^{5}$ & $1.8\times10^{7}$ & $1.7\times10^{7}$ & $3.5\times10^{7}$ & $1.6\times10^{9}$ & $1.6\times10^{9}$ & $3.2\times10^{9}$ \\
        supeconductor & $\mathrm{K_3C_{60}}$ & $5.0\times10^{5}$ & $4.2\times10^{5}$ & $9.2\times10^{5}$ & $3.7\times10^{7}$ & $3.7\times10^{7}$ & $7.4\times10^{7}$ & $3.6\times10^{9}$ & $3.6\times10^{9}$ & $7.2\times10^{9}$ \\ \hline
        \multirow{4}{*}{} & LaFeAsO & $6.5\times10^{6}$ & $5.2\times10^{6}$ & $1.2\times10^{7}$ & $4.2\times10^{8}$ & $4.1\times10^{8}$ & $8.4\times10^{8}$ & $4.0\times10^{10}$ & $4.0\times10^{10}$ & $8.0\times10^{10 }$\\ 
        Iron-based & $\mathrm{BaFe_2As_2}$ & $6.7\times10^{6}$ & $5.2\times10^{6}$ & $1.2\times10^{7}$ & $4.3\times10^{8}$ & $4.1\times10^{8}$ & $8.4\times10^{8}$ & $4.0\times10^{10}$ & $4.0\times10^{10}$ & $8.0\times10^{10 }$\\ 
        superconductor & LiFeAs & $1.3\times10^{7}$ & $8.3\times10^{6}$ & $2.1\times10^{7}$ & $4.9\times10^{8}$ & $4.4\times10^{8}$ & $9.3\times10^{8}$ & $4.1\times10^{10}$ & $4.0\times10^{10}$ & $8.1\times10^{10} $\\ 
        ~ & FeSe & $1.7\times10^{7}$ & $1.0\times10^{7}$ & $2.8\times10^{7}$ & $5.3\times10^{8}$ & $4.6\times10^{8}$ & $1.0\times10^{9}$ & $4.1\times10^{10}$ & $4.1\times10^{10}$ & $8.2\times10^{10} $\\ \hline
        \multirow{4}{*}{} & MnO & $1.8\times10^{6}$ & $1.4\times10^{6}$ & $3.2\times10^{6}$ & $1.1\times10^{8}$ & $1.0\times10^{8}$ & $2.1\times10^{8}$ & $1.0\times10^{10}$ & $1.0\times10^{10}$ & $2.0\times10^{10} $\\ 
        Binary transition & FeO & $3.0\times10^{6}$ & $2.0\times10^{6}$ & $5.1\times10^{6}$ & $1.2\times10^{8}$ & $1.1\times10^{8}$ & $2.3\times10^{8}$ & $1.0\times10^{10}$ & $1.0\times10^{10}$ & $2.0\times10^{10} $\\ 
        metal oxide & CoO & $7.9\times10^{6}$ & $4.4\times10^{6}$ & $1.2\times10^{7}$ & $1.7\times10^{8}$ & $1.3\times10^{8}$ & $3.0\times10^{8}$ & $1.1\times10^{10}$ & $1.0\times10^{10}$ & $2.1\times10^{10} $\\ 
        ~ & NiO & $9.5\times10^{6}$ & $5.2\times10^{6}$ & $1.5\times10^{7}$ & $1.9\times10^{8}$ & $1.4\times10^{8}$ & $3.3\times10^{8}$ & $1.1\times10^{10}$ & $1.0\times10^{10}$ & $2.1\times10^{10} $\\ \hline
        \multirow{3}{*}{Perovskite oxide} & $\mathrm{SrVO_3}$ & $2.9\times10^{6}$ & $1.9\times10^{6}$ & $4.8\times10^{6}$ & $1.2\times10^{8}$ & $1.1\times10^{8}$ & $2.3\times10^{8}$ & $1.0\times10^{10}$ & $1.0\times10^{10}$ & $2.0\times10^{10} $\\ 
        ~ & $\mathrm{SrCrO_3}$ & $2.5\times10^{6}$ & $1.7\times10^{6}$ & $4.2\times10^{6}$ & $1.1\times10^{8}$ & $1.1\times10^{8}$ & $2.2\times10^{8}$ & $1.0\times10^{10}$ & $1.0\times10^{10}$ & $2.0\times10^{10} $\\ 
        ~ & $\mathrm{SrMnO_3}$ & $2.9\times10^{6}$ & $2.0\times10^{6}$ & $4.9\times10^{6}$ & $1.2\times10^{8}$ & $1.1\times10^{8}$ & $2.3\times10^{8}$ & $1.0\times10^{10}$ & $1.0\times10^{10}$ & $2.0\times10^{10} $\\ \hline \hline
        \multicolumn{2}{c}{Average} & $5.7\times10^{6}$ & $3.7\times10^{6}$ & $9.5\times10^{6}$ & $2.2\times10^{8}$ & $2.0\times10^{8}$ & $4.2\times10^{8}$ & $1.8\times10^{10}$ & $1.8\times10^{10}$ & $3.7\times10^{10} $\\ \hline
    \end{tabularx}
\end{center}
\end{table*}

\subsection{Results of the gate resource estimation}
\label{sec:Gate resource estimation results}
Table~\ref{tab:List of the resource estimation results} shows the number of gates $N_{gates}[N_{cells}]$ calculated for the 13 target compounds.
For example, the values of $N_{gates}[N_{cells}]$ for $10^2$ of $N_{cells}$ range from $10^5$ to $10^7$ for the one-qubit gate, CNOT gate, and the summation. The orders of $N_{gates}$ for the one-qubit gate, CNOT gate, and summation are almost the same in each compound and in each cell size. 
Since the estimated values for resources are much larger than the number of gates performed in the current quantum devices $\sim10^3$~\cite{Preskill2018-sc,Huang2022-ke}, then gate resource reduction is necessary to perform the Hamiltonian simulation.
Now we discuss only the values of the summation, ``Sum,'' in Table~\ref{tab:List of the resource estimation results}.

\begin{figure}[!b]
 \includegraphics[width=1\columnwidth]{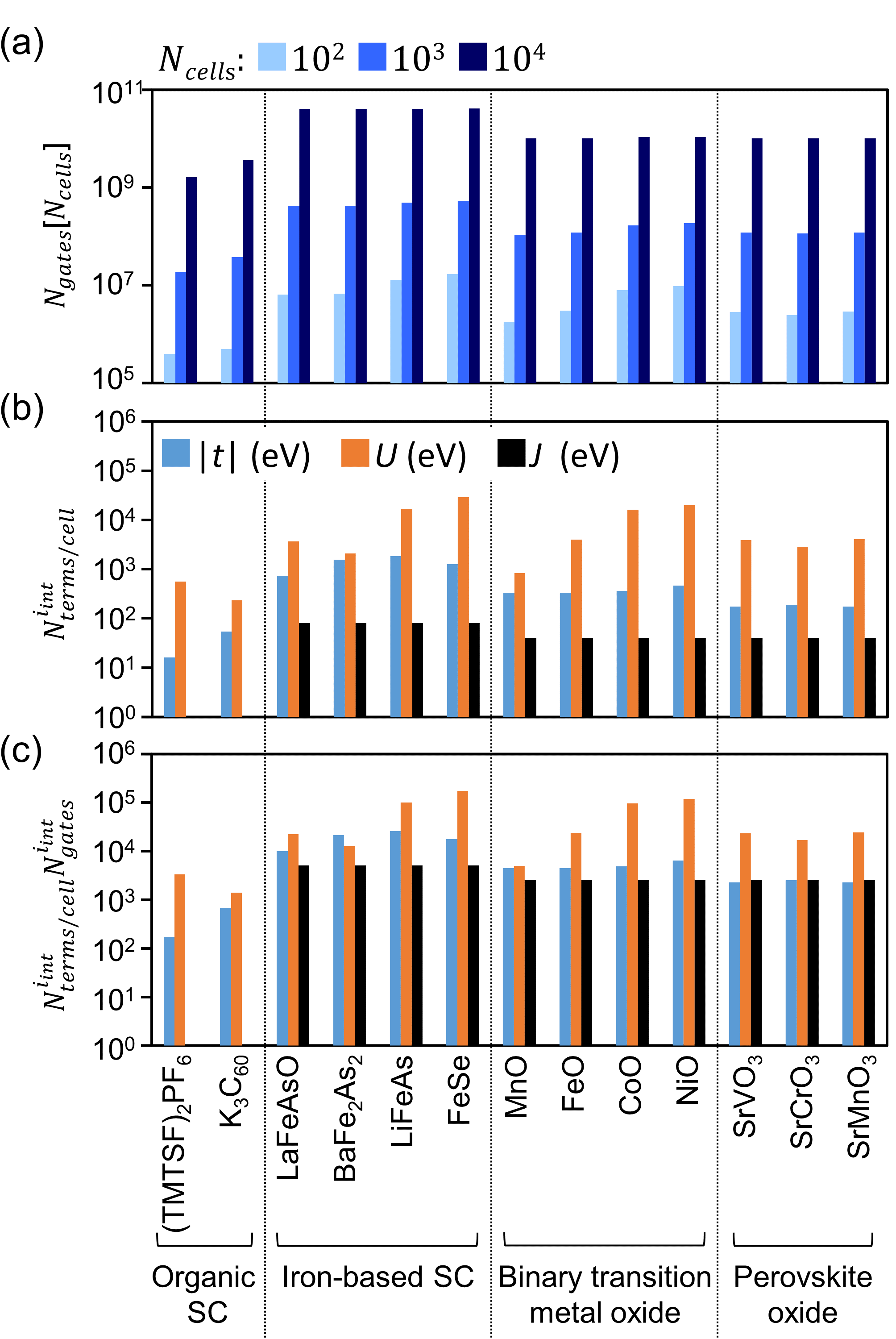}
\caption{Resource estimation results for the number of gates for target compounds. (a) The number of gates $N_{gates}[N_{cells}]$ for $10^2, 10^3$, and $10^4$ of $N_{cells}$ . (b) The number of interaction terms per unit cell $N_{terms/cell}^{i_{int}}$. (c) The number of gates of the interaction term per unit cell $N_{terms/cell}^{i_{int}} N_{gates}^{i_{int}}$. 
The values of (b) and (c) are shown for $|t|$, $U$, and $J$.}
 \label{fig:Estimation_results_1col}
\end{figure}

\begin{figure}[!h]
 \includegraphics[width=1\columnwidth]{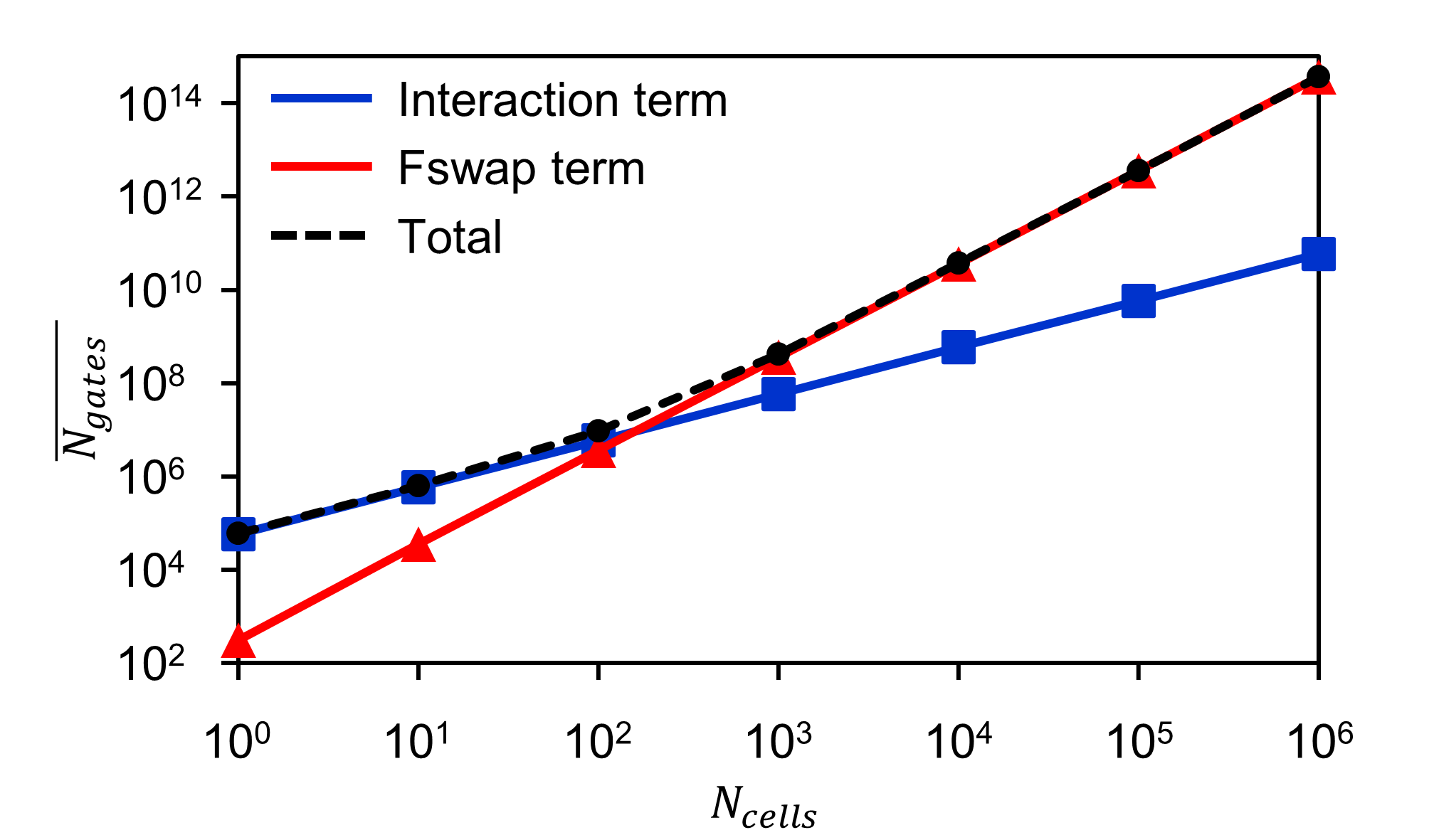}
\caption{$N_{cells}$ dependence for the number of gates in the interaction term (blue line), fswap term (red line), and the total (black dash line) in Eq.~(\ref{eq:Ngates(Ncell)}). The values are averaged over all the compounds.}
 \label{fig:Comparison_fswap_interaction_1col}
\end{figure}

Figure~\ref{fig:Estimation_results_1col}(a) shows the values of $N_{gates}[N_{cells}]$ for ``Sum'' in Table~\ref{tab:List of the resource estimation results}.
The results show that the larger $N_{cells}$, the weaker the compound dependence on $N_{gates}[N_{cells}]$ in the same class of the compounds.
For example, in the binary transition metal oxide, the ratio of the value for NiO to that for MnO is smaller in $10^4$ of $N_{cells}$ than in $10^2$ of $N_{cells}$.
This is because the values of the interaction (first) term and fswap (second) term in Eq.~(\ref{eq:Ngates(Ncell)}) are $O(N_{cells})$ and $O(N_{cells}^2)$, respectively.
In other words, as $N_{cells}$ becomes larger, the fswap operation becomes more dominant than that for the interaction operation.
Figure~\ref{fig:Comparison_fswap_interaction_1col} shows the number of gates for the interaction and fswap terms as a function of $N_{cells}$, where the values are averaged over all the compounds.
When $N_{cells}$ increases from $10^2$ to $10^3$, the value of the fswap term exceeds that of the interaction term.
We mention the comparison of the results of the simple model such as the Hubbard model with the present results.
In the resource estimation for the Hubbard model~\cite{Cai2020-sv}, the number of gates required for simulating a single Trotter step at 25 sites ($=50$ qubits) is $10^3$, whereas in the present correlated electron materials, the number of gates required for a similar number of qubits ($N_{cells}\sim5$, e.g., 5 sites in NiO) is about $10^5$. Thus, the required gate resources for simulations of correlated electron materials would be much larger by several orders than simple model simulations of the same size.

Next, we analyze the influence of the interaction operations on the gate resources for relatively small systems (i.e., $N_{cells} \lesssim10^2$). 
We show the number of interaction terms per unit cell, $N_{terms/cell}^{i_{int}}$, in Fig.~\ref{fig:Estimation_results_1col}(b), and the number of gates of the interaction term per unit cell, $N_{terms/cell}^{i_{int}} N_{gates}^{i_{int}}$, in Fig.~\ref{fig:Estimation_results_1col}(c).
Here, $N_{terms/cell}^{i_{int}}$ for $|t|$ denotes the summation of the number of interaction terms for $t_{pp\sigma}$ and $t_{pq\sigma}$; $N_{terms/cell}^{i_{int}}$ for $U$ denotes the summation for $U_{pp}$ and $U_{pq}$; $N_{terms/cell}^{i_{int}}$ for $J$ denotes that for $J_{pq}$.
$N_{gates}^{i_{int}}$ for $|t|$, $U$, and $J$ are 15, 12, and 246, respectively, which are the total values in the row 1-2, 3-4, and 5-7 in the column ``Sum'' in Table~\ref{tab:the number of gates}.
Figure~\ref{fig:Estimation_results_1col}(b) indicates that $N_{terms/cell}^{i_{int}}$ for $U$ is larger than those for $|t|$ and $J$ for all the compounds, and except for $\mathrm{BaFe_2As_2}$, Fig.~\ref{fig:Estimation_results_1col}(c) holds the same trend. 
Note that since $N_{terms/cell}^{i_{int}}$ for $U$ and $|t|$ of $\mathrm{BaFe_2As_2}$ are close and $N_{gates}^{i_{int}}$ for $|t|$ is larger than that for $U$, $N_{terms/cell}^{i_{int}}N_{gates}$ for $U$ of $\mathrm{BaFe_2As_2}$ is smaller than that for $|t|$.
Furthermore, there are no terms for $J$ to be counted for $N_{terms/cell}^{i_{int}}$ in the class of the organic superconductors.
As a result, we found that constructing an effective Hamiltonian that reduces the number of terms related to the Coulomb interaction is effective in reducing resources for relatively small systems.

For relatively large systems (i.e., $N_{cells} \gtrsim 10^3$), reducing the gate resource related to the fswap operations is effective. 
Especially, reducing $N_{pairswap}$ in Eq.~(\ref{eq:Ngates}), i.e., the number of the pair swapping or fswap operations has a larger effect on the resource than reducing $N_{gates\mathchar`-fswap}$ since the value of $N_{gates\mathchar`-fswap}$ is only four.
Therefore, it is worth considering the elimination of unnecessary fswap operations and the search for an efficient swapping order.
Besides, the parallel execution of the fswap operation reduces the depth in the circuit even with the same number of gates as the serial execution.
Specifically, the depth for the fswap operation reduces from $4\times{\frac{N_{qubits/cell} N_{cells}}{2} \choose 2}$ to $3\times\frac{N_{qubits/cell} N_{cells}}{2}$ using the parallel execution.

\section{Conclusion}
\label{sec:Conclusion}
In this study, we estimated the resources required for the Hamiltonian simulation in electron correlated materials. Specifically, we estimated the number of quantum gates and qubits required for the Trotter-based Hamiltonian simulation (per Trotter step) using the fermionic swap (fswap) network.
Here, 13 target compounds classified as organic superconductors, iron-based superconductors, binary transition metal oxides, and perovskite oxides were selected for the resource estimation.
We adopted the effective Hamiltonian with the orbitals around the Fermi level using the $ab~initio$ downfolding method that uses density functional theory, maximally localized Wannier function, and constrained random phase approximation.
Furthermore, we proposed the procedure that swaps any two pairs of up and down spin to perform the fswap network on the effective Hamiltonian with the exchange interactions.

We obtained that the estimated values of the numbers of the gates per Trotter step and qubits are $10^7$ ($10^8$) and $10^3$ ($10^4$), respectively, on averages for the 13 compounds with $10^2$ ($10^3$) of $N_{cells}$, where $N_{cells}$ denotes the number of unit cells in the system.
For example, $1.2\times10^7$ ($8.4\times10^8$) gates per Trotter step and $2\times10^3$ ($2\times10^4$) qubits are required in the system for $10^2$ ($10^3$) of $N_{cells}$ in LaFeAsO.

Moreover, we analyzed the gate resource and found that the gate resources for the interaction and fswap operations scale as $O(N_{cells})$ and $O(N_{cells}^2)$, respectively.
Additionally, on the average value of the 13 compounds, the number of interaction terms in the effective Hamiltonian, especially that of the Coulomb interaction terms, is dominant to the gate resource up to $N_{cells}\sim10^2$.
Therefore, reducing the number of Coulomb interaction terms is effective for gate cost reduction in relatively small systems.
In relatively large systems ($N_{cells}\gtrsim 10^3$), the resource for the fswap operations, especially the number of fswap operations, is dominant, and thus reducing the number of fswap operations is effective.

Additionally, the error mitigation technique~\cite{Endo2021-ku,Takagi2021-ne,Temme2017-vo,Endo2018-zg,Takagi2021-kt,Endo2019-oh} may be useful because it can increase the implementable number of gates in a quantum circuit when the number of errors in a quantum circuit is in the order of unity.
Using the probabilistic error cancellation~\cite{Temme2017-vo,Endo2018-zg}, a reliable quantum simulation can be realized when the condition $\varepsilon_{TG}N_{TG}\lesssim2$ is kept~\cite{Endo2020-ct}, where $\varepsilon_{TG}$ is the error rate of two-qubit gates, and $N_{TG}$ is the number of two-qubit gates. The condition is derived from the fact that the two-qubit gate can be regarded as a main noise source of a quantum device. When $N_{cells}$ is $10^2$, i.e., $10^3$ qubits, the number of the two-qubit (CNOT) gates required for the Hamiltonian simulation of $N_{steps}$ steps is $N_{TG}\sim N_{steps}\times10^6$ (see “Average” row in Table~\ref{tab:List of the resource estimation results}). Therefore, when a quantum device can realize an error rate of $\varepsilon_{TG}\lesssim 2 N_{steps}^{-1}\times10^{-6}$, the number of gates required for performing the Hamiltonian simulation in the size of $N_{cells}=10^2$ will be executable (for example, see Refs.~\cite{Lloyd1996-wk,Reiher2017-jt} for the value of $N_{steps}$).

\section{Acknowledgments}
This work was partly supported by JSPS KAKENHI Grant Number 21H01742.

\bibliographystyle{apsrev4-1}
\bibliography{bib_esitmation}

\appendix
\section{The quantum circuit implementations in the fswap network}
\label{The quantum circuit implementations in the fswap network}
\begin{figure}[!h]
 \includegraphics[width=1\columnwidth]{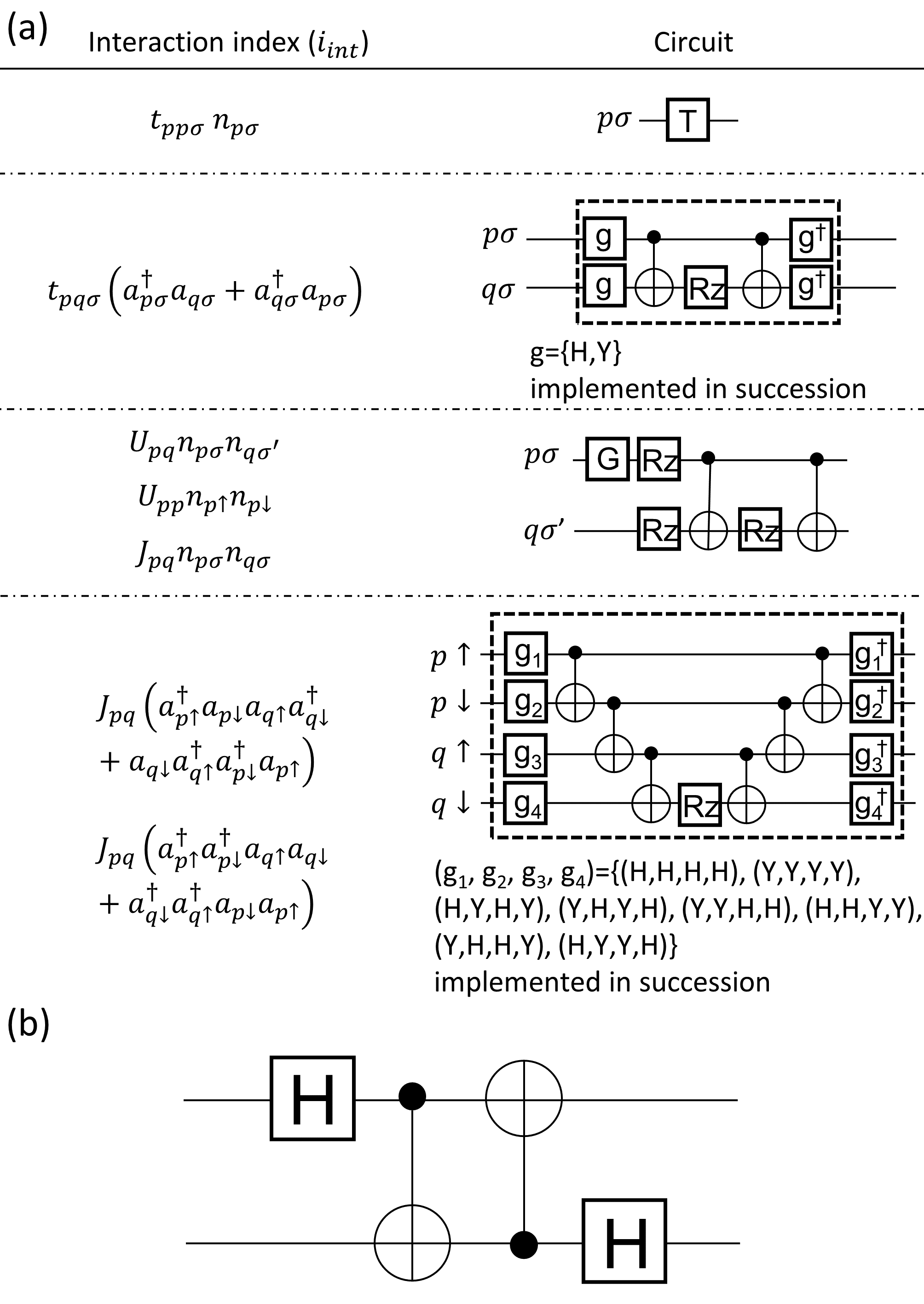}
\caption{Quantum gates corresponding to each interaction operator of $i_{int}$ and the fswap operator in our procedure. 
$T$ in the box of the circuit is the one-qubit gate as
$\begin{pmatrix}
1 & 0 \\
0 & e^{-i\theta} \\
\end{pmatrix}$
, $G$ is the global phase gate as
$e^{-i\theta}
\begin{pmatrix}
1 & 0 \\
0 & 1 \\
\end{pmatrix}$
, $H$ is the Hadamard gate, $Y$ is the $R_y(-\frac{\pi}{2})$ gate, and $R_z$ is the $R_z(\theta)$ gate, where $\theta$ is the parameter determined with the interaction coefficient (see Ref.~\cite{Whitfield2011-lz} for details).
(a) Quantum gates for each interaction operator of $i_{int}$. 
The spin-orbital indices are indicated on the left side of each circuit.
The part enclosed by a dashed line in the circuit of the second row is assigned to a circuit with $g=H$ substituted, followed by a circuit with $g=Y$ in succession.
The similar operation is performed for \{$g_1, g_2, g_3, g_4$\} in the last row.
(b) A quantum gate for the fswap operator between the neighboring qubits.}
 \label{fig:circuit_list_1col}
\end{figure}

The number of gates in the interaction operations in Table~\ref{tab:the number of gates} and that in the fswap operation are derived from the circuits in Figs.~\ref{fig:circuit_list_1col}(a) and (b), respectively.
We derived the circuit implementations from the descriptions of Ref.~\cite{Whitfield2011-lz} in Fig.~\ref{fig:circuit_list_1col}(a) and Refs.~\cite{Cai2020-sv,Schuch2003-zq} in Fig.~\ref{fig:circuit_list_1col}(b).
In Fig.~\ref{fig:circuit_list_1col}(a), $i_{int}$ is the interaction index, i.e., one of the seven interactions in Eq.~(\ref{eq:effective H}) (e.g., $t_{p p \sigma} n_{p \sigma}$).

In our procedure of the fswap network, the spin-orbital indices mapped to the qubits are changed at each step in the procedure; the initial sequence of the indices is $(1\uparrow, 1\downarrow, 2\uparrow, 2\downarrow, \dots)$, and the neighboring indices are swapped according to the procedure in Fig.~\ref{fig:fswap_procedure_1col} and Algorithm~\ref{Alg:fswap} by using the fswap operations.
The sequence of the spin-orbital indices shown in Fig.~\ref{fig:circuit_list_1col}(a) can appear during the fswap operations, in which two qubits connected with operations are physically neighboring pairs.
Figure ~\ref{fig:circuit_list_1col}(b) shows the circuit for the fswap operation on neighboring qubits.

\end{document}